\documentclass[a4paper,twocolumn,11pt,accepted=2023-01-04]{quantumarticle}
\pdfoutput=1
\usepackage[utf8]{inputenc}
\usepackage[english]{babel}
\usepackage[T1]{fontenc}
\usepackage{amsmath}
\usepackage{hyperref}
\usepackage{graphicx}
\usepackage[normalem]{ulem}
\usepackage{dcolumn}
\usepackage{epsfig}
\usepackage{bm}
\usepackage{array}
\usepackage{lipsum} 
\usepackage{tikz}
\usetikzlibrary{calc,arrows,chains,matrix,positioning,scopes}
\usepackage[numbers]{natbib}
\usepackage{placeins}
\usepackage{float}

\hypersetup{
colorlinks=true,
citecolor=blue,
linkcolor=red,
urlcolor=blue
,pdfmenubar=true
}

\usepackage{amsfonts}
\usepackage{amssymb}
\usepackage{xcolor}
\usepackage{bbm}
\usepackage{calrsfs}
\usepackage{dutchcal}
\usepackage{amsthm}

\usepackage{txfonts}

\AtBeginDocument{%
    \newwrite\bibnotes
    \def\bibnotesext{Notes.bib}
    \immediate\openout\bibnotes=\jobname\bibnotesext
    \immediate\write\bibnotes{@CONTROL{REVTEX41Control}}
    \immediate\write\bibnotes{@CONTROL{%
    apsrev41Control,author="08",editor="1",pages="1",title="0",year="1"}}
     \if@filesw
     \immediate\write\@auxout{\string\citation{apsrev41Control}}%
    \fi
}%

\newcommand{\new}[1]{\textcolor{black}{#1}}

\newcommand{\be}{\begin{equation}}
\newcommand{\ee}{\end{equation}}

\newcommand{\beq}{\begin{eqnarray}}
\newcommand{\eeq}{\end{eqnarray}}

\begin{document}

\title{Extending the fair sampling assumption using causal diagrams}
\author{Valentin~Gebhart}
\affiliation{QSTAR, INO-CNR and LENS, Largo Enrico Fermi 2, 50125 Firenze, Italy}
\email{gebhart@lens.unifi.it}

\author{Augusto~Smerzi}
\affiliation{QSTAR, INO-CNR and LENS, Largo Enrico Fermi 2, 50125 Firenze, Italy}
\email{augusto.smerzi@ino.cnr.it}

\begin{abstract}
Discarding undesirable measurement results in Bell experiments opens the detection loophole that prevents a conclusive demonstration of nonlocality. As closing the detection loophole represents a major technical challenge for many practical Bell experiments, it is customary to assume the so-called fair sampling assumption (FSA) that, in its original form, states that the collectively postselected statistics are a fair sample of the ideal statistics. Here, we analyze the FSA from the viewpoint of causal inference: \new{We derive a causal structure that must be present in any causal model that faithfully encapsulates the FSA.} This provides an easy, intuitive, and unifying approach that includes different accepted forms of the FSA and underlines what is really assumed when using the FSA. We then show that the FSA can not only be applied in scenarios with non-ideal detectors or transmission losses, but also in ideal experiments where only parts of the correlations are postselected, e.g., when the particles' destinations are in a superposition state. Finally, we demonstrate that the FSA is also applicable in multipartite scenarios that test for (genuine) multipartite nonlocality. 
\end{abstract}

\maketitle

\section{Introduction}\label{sec:introduction}

Bell nonlocality~\cite{bell1964,bell1976} represents one of the central pillars of modern research in quantum foundations and the development of quantum technologies~\cite{brunner2014}. 
A widely-used technique in Bell experiments is the discard of events of an incomplete detection such as, e.g., the non-detection of parts of the system due to particle losses. 
By the selection bias~\cite{pearl2009}, the postselection opens the detection loophole, i.e., the possibility of a local hidden variable (LHV) model to describe the observed correlations even if the postselected correlations violate a Bell inequality~\cite{pearle1970,clauser1974}.
The detection loophole is not just conspiratorial: 
It was used in experiments to create fake violations of Bell inequalities~\cite{tasca2009,gerhardt2011,pomarico2011,romero2013}. 

Ideally, one can close the detection loophole by including the non-detection events~\cite{clauser1974,mermin1986,eberhard1993,sciarrino2011}, or by sharpening the Bell inequalities~\cite{garg1987,larsson1998,gebhart2022}. 
These methods require high detection efficiencies and have recently been implemented in sophisticated Bell experiments that close the detection loophole~\cite{rowe2001,matsukevich2008,christensen2013} (also while simultaneously closing the locality loophole~\cite{shalm2015,giustina2015,hensen2015}).
However, the required detection efficiencies represent a severe technical challenge for the practicality of Bell experiments. 
Thus, a widely-used way out is to rely on the fair sampling assumption (FSA)~\cite{clauser1969,clauser1974,berry2010,orsucci2020} that is commonly known as the assumption that the postselected statistics is a fair representation of (i.e., is identical to) the statistics that would have been observed using perfect detectors and no losses. 
An alternative form of the FSA is the assumption that the detector settings have no influence on the detection probability of the particles. 
In the latter case, the postselected statistics need not be identical to the statistics that would be observed with ideal detectors but, nonetheless, the postselection cannot create any fake nonlocal correlations. 
We emphasize that the FSA does not close the detection loophole, but it rather represents an assumption that restricts the possible LHV descriptions for the measurement data.
Due to the high technical requirements of closing the detection loophole (e.g., highly efficient detectors and minimal transmission losses), the FSA is still widely used in Bell experiments~\cite{marinkovic2018,rauch2018,polino2019,gomez2019,poderini2020,velez2020,agresti2020}. 

In this work, we analyze the FSA from the viewpoint of causal inference and causal diagrams~\cite{pearl2009}. 
In particular, we ask what structure any causal diagram of a Bell experiment must possess to allow for a valid demonstration of nonlocality if the data is collectively postselected.
This structure should guarantee that the postselection cannot create fake nonlocal correlations due to the selection bias. Importantly, we ensure that the causal diagram provides a meaningful description of the experiment by disallowing any kind of fine-tuning of causal influences~\cite{pearl2009,spirtes2000,wood2015}, in contrast to previous studies of the FSA. This results in an easy and intuitive way to understand different forms of the FSA found in the literature. Our analysis highlights what is really assumed when using the FSA, and allows us to identify Bell scenarios where no faithful causal explanation of the FSA exists. Furthermore, we show that the such-obtained causal-diagram FSA can be applied to different experiments where the correlations must be postselected even in the ideal noiseless setup, and not just in the standard setup with non-ideal detectors and particle losses. Finally, we prove that the FSA is also useful in experiments that demonstrate multipartite nonlocality and genuine multipartite nonlocality.

\section{Fair sampling in the bipartite scenario}\label{sec:bipartite}

In the following, we derive a necessary causal structure for any faithful causal model of a bipartite Bell scenario that includes a collective postselection, without potentially creating fake correlations that violate the Bell inequality. The central promise is to have a causal description that does not employ any fine-tuning of its causal influences. By a fine-tuning of the parameters of a causal model, two variables can be made statistically independent even though they seem to affect each other in the causal diagram. 
Instead, without fine-tuning, any statistical independence between two variables must be evident from the diagram's structure. If one allows for fine-tuning, the description in terms of causal diagrams becomes irrelevant~\cite{pearl2009,spirtes2000,wood2015} because any statistical (in)dependence can just be directly imposed by hand. 
Causal models that are not fine-tuned are usually called faithful models.

The central tool to infer independencies from a causal diagram are the $d$-separation rules~\cite{pearl2009} that dictate whether a given path connecting two variables of the causal diagram is open (i.e., the variables are generally dependent) or blocked (the variables must be independent), also when conditioning on other variables of the diagram. 
The rules state that {(i)} a path is blocked if along it there is a collider (a variable where two causal arrows collide), ({ii}) a path is blocked if along it there is a non-collider that is conditioned on, and ({iii}) a path is open if along it there is a collider and we condition on the collider or its descendants. The last rule manifests itself in the selection bias and the Berkson paradox~\cite{pearl2009}. 
We want to note that in the context of Bell experiments, causal diagrams and the $d$-separation rules have been used to show that quantum violations of Bell inequalities require fine-tuning in classical-causal explanations (e.g., superdeterminism, superluminal influences, or retrocausal influences)~\cite{wood2015,allen2017,cavalcanti2018}, and to show that certain collective postselection strategies are safe for the demonstration of multipartite nonlocality~\cite{blasiak2021} and genuine multipartite nonlocality~\cite{gebhart2021,gebhart2022}. In the latter works, it was shown that if the postselection can be decided by excluding some of the parties, the detection loophole can be closed, so one does not have to rely on the FSA. In contrast, here we consider a postselection that must be decided by all parties together, such that the results of Refs.~\cite{blasiak2021,gebhart2021,gebhart2022} do not apply.

\begin{figure}[t]
\centering
\begin{minipage}{.49\linewidth}
\begin{tikzpicture}[thick, every node/.style={scale=1.2}]
\node[blue!70!black,text height=7pt,text depth=2pt] (lambda) {$\Lambda$};
\node[text height=7pt,text depth=2pt,below left = .8cm and .0cm of lambda] (A) {$A$};
\node[text height=7pt,text depth=2pt,below right = .8cm and .0cm of lambda] (B) {$B$};
\node[text height=7pt,text depth=2pt,above left = .3cm and .0cm of A] (X) {$X$};
\node[text height=7pt,text depth=2pt,above right = .3cm and .0cm of B] (Y) {$Y$};
\node[text height=7pt,text depth=2pt,below = 1.8cm of lambda] (K) {$\quad$};
\node[text height=7pt,text depth=2pt,above left = .2cm and -.5cm of X] (label) {\footnotesize \textbf{(a)}};
\node[text height=7pt,text depth=2pt,below = 2.0cm of lambda] (K) {};

\draw[blue!70!black,->] (lambda) edge (A)
          (lambda) edge (B)
;
\draw[->]
          (X) edge (A)
          (Y) edge (B)
;          
\end{tikzpicture}
\end{minipage}
\begin{minipage}{.49\linewidth}
\begin{tikzpicture}[thick, every node/.style={scale=1.2}]
\node[blue!70!black,text height=7pt,text depth=2pt] (lambda) {$\Lambda$};
\node[text height=7pt,text depth=2pt,below left = .8cm and .0cm of lambda] (A) {$A$};
\node[text height=7pt,text depth=2pt,below right = .8cm and .0cm of lambda] (B) {$B$};
\node[text height=7pt,text depth=2pt,above left = .3cm and .0cm of A] (X) {$X$};
\node[text height=7pt,text depth=2pt,above right = .3cm and .0cm of B] (Y) {$Y$};
\node[draw,rectangle,red!70!black,text height=7pt,text depth=2pt,below = 2.0cm of lambda] (K) {$K$};
\node[text height=7pt,text depth=2pt,above left = .2cm and -.5cm of X] (label) {\footnotesize \textbf{(b)}};

\draw[blue!70!black,->] (lambda) edge (A)
          (lambda) edge (B)
;
\draw[->]
          (X) edge (A)
          (Y) edge (B)
;          
\draw[->,red!70!black] 
          (A) edge (K)
          (B) edge (K)
;

\end{tikzpicture}
\end{minipage}
\caption{(a) The causal diagram of the local hidden variable (LHV) model in the standard bipartite Bell scenario. (b) A collective postselection is indicated as the variable $K$ that is influenced by both Alice's outcome $A$ and Bob's outcome $B$, opening the detection loophole by the selection bias~\cite{pearle1970}. The conditioning on the decision of the postselection is indicated as a box around $K$.}
\label{fig:LHVbipartite}
\end{figure}
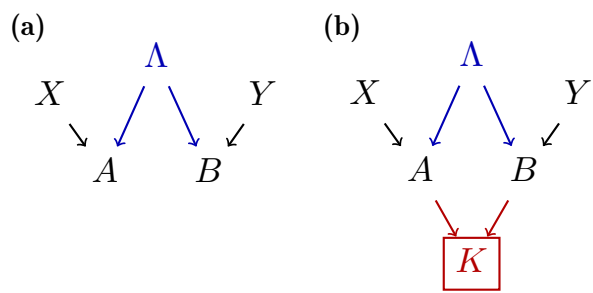

In the standard bipartite Bell scenario, two parties, Alice and Bob, share the two parts of a quantum system and each perform local measurements on their subsystem. Alice (Bob) chooses the measurement setting $x$ ($y$) and records the measurement outcome $a$ ($b$). The measured correlations are called local if they can be described by a LHV model of the form~\cite{bell1964,bell1976}
\begin{equation}\label{eq:LHV}
    p_{ab|xy}=\sum_\lambda p_\lambda p_{a|x\lambda} p_{b|y\lambda},
\end{equation}
where $p_\lambda$, $p_{a|x\lambda}$, and $p_{b|y\lambda}$ are (conditional) probabilities, each summing to $1$, e.g., $\sum_\lambda p_\lambda=1$.
The corresponding causal diagram is shown in Fig.~\ref{fig:LHVbipartite}(a). Using Eq.~\eqref{eq:LHV}, one can derive Bell inequalities, a violation of which proves that the correlations are nonlocal.

In Fig.~\ref{fig:LHVbipartite}(b), we include the variable $K$ representing the decision of the collective postselection (e.g., $K=1$ for postselecting the results and $K=0$ for discarding the results). If the postselected statistics $p_{ab|xy1}$ ($1$ is the value of the variable $K$) can be described by a LHV similar to Eq.~\eqref{eq:LHV}, they must also fullfill the Bell inequality and the postselection is valid. By the definition of conditional probability, we can write $p_{ab|xyk}=\sum_\lambda p_{\lambda|xyk} p_{ab|xy\lambda k}$, such that we can identify the two conditions
\begin{align*}
    \mathrm{\textbf{CI}:} & \quad \quad \quad  p_{\lambda | x y k} = p_{\lambda | k}\\
    \mathrm{\textbf{CII}:} & \quad \quad p_{ab|xy\lambda k} = p_{a|x\lambda k}p_{b|y\lambda k}
\end{align*}
that ensure a LHV description of $p_{ab|xyk}$ and thus a valid postselection\footnote{It would actually suffice to show that $p_{ab|xyk}$ admits a LHV model only for $k=1$, and not necessarily for all $k$. The causal-inference techniques that we employ below yield the conditions for all $k$, implying the case $k=1$.}. We note that \textbf{CI} and \textbf{CII} correspond to the measurement-independence and locality assumptions of Bell's theorem, respectively. 
Now consider a postselection that is collective: Both parties must consult each other to decide the postselection. Note that if each party can decide the postselection locally, there is no need for a FSA because the postselection is known to be safe~\cite{sciarrino2011,blasiak2021}.
Therefore, in the causal diagram, $K$ is influenced by both measurement outcomes $A$ and $B$~\footnote{\label{footnote1}Strictly speaking, a collective postselection could also include direct influences from the settings $X$ and $Y$ to $K$, e.g., a postselection influenced only by $X$ and $B$, or only by $X$ and $Y$. A postselection influenced by $X$ and $B$ leads to a conflict with condition $\textbf{CI}$, similar a postselection influenced by $A$ and $B$ as in the main text. On the other side, a postselection that is decided only by $X$ and $Y$ is a safe postselection, i.e., the conditions $\textbf{CI}$ and $\textbf{CII}$ still hold. This can be seen by noting that, if $K$ is influenced only by $X$ and $Y$, one simply has $p_{ab|xyk}=p_{ab|xy}$.}, see Fig.~\ref{fig:LHVbipartite}(b). 
However, a postselection described by the causal diagram in Fig.~\ref{fig:LHVbipartite}(b) is in conflict with condition $\textbf{CI}$: There is an open path $X\rightarrow A \rightarrow K \leftarrow B \leftarrow \Lambda$ ($K$ is a collider that is conditioned on), which contradicts $p_{\lambda|xyk}=p_{\lambda|k}$.

The causal diagram of Fig.~\ref{fig:LHVbipartite}(b) can thus not give a causal account of the fair sampling without employing a fine-tuning of causal influences. 
In the following, we show that any causal description of the FSA requires a certain type of structure in the causal model if the model is not fine-tuned.
We consider a general bipartite Bell scenario where Alice's (Bob's) measurement settings are given by a number of setting variables $\mathbf{x}=(x_1,\dots,x_{n_A})$ [$\mathbf{y}=(y_1,\dots,y_{n_B})$], and Alice (Bob) observes a number of measurement outcomes $\mathbf{a}=(a_1,\dots,a_{m_A})$ [$\mathbf{b}=(b_1,\dots,b_{m_B})$]. Their measurement outcomes are correlated via the LHV $\Lambda$ (note that we group all LHVs into the single LHV $\Lambda$ without loss of generality). At this point, the different outcome variables of each party can be arbitrarily connected by causal influences, e.g., one could have a causal influence $A_1\rightarrow A_2$. After performing their measurements, the parties collectively postselect their data, represented as the binary postselection variable $K$ as above. The corresponding causal structure is identical to the one of Fig.~\ref{fig:LHVbipartite}(b), expect that all setting and outcome variables are replaced by multi-variable versions. Thus, similar to above, this general causal model is in conflict with the conditions $\textbf{CI}$ and $\textbf{CII}$ for a valid postselection.

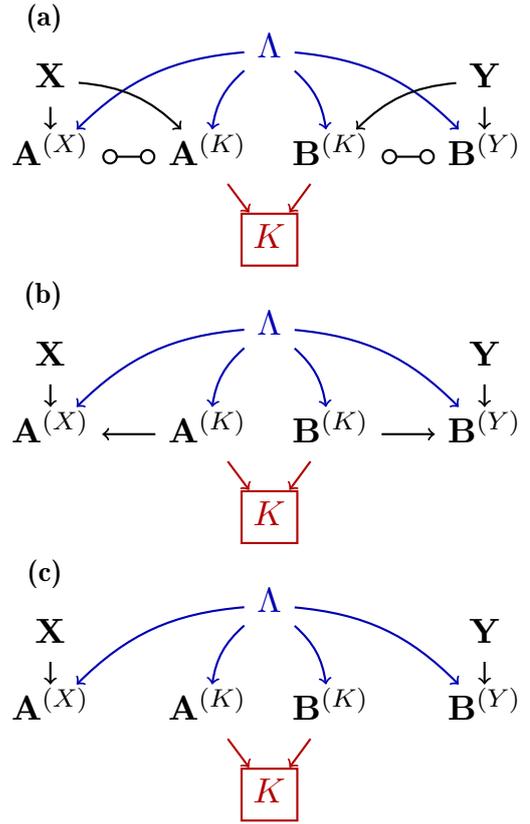
\begin{figure}[t]
\centering
\begin{tikzpicture}[thick, every node/.style={scale=1.2}]
\node[blue!70!black,text height=7pt,text depth=2pt] (lambda) { $\Lambda$};
\node[text height=7pt,text depth=2pt,below left = .7cm and -.2cm of lambda] (A2) {$\mathbf{A}^{(K)}$};
\node[text height=7pt,text depth=2pt, left = .7cm of A2] (A1) {$\mathbf{A}^{(X)}$};
\node[text height=7pt,text depth=2pt,below right = .7cm and -.2cm of lambda] (B2) {$\mathbf{B}^{(K)}$};
\node[text height=7pt,text depth=2pt, right = .7cm of B2] (B1) {$\mathbf{B}^{(Y)}$};
\node[text height=7pt,text depth=2pt,above = .3cm of A1] (X) {$\mathbf{X}$};
\node[text height=7pt,text depth=2pt,above = .3cm of B1] (Y) {$\mathbf{Y}$};
\node[draw,rectangle,red!70!black,text height=7pt,text depth=2pt,below = 1.8cm of lambda] (K) {$K$};
\node[text height=7pt,text depth=2pt,above left = 0.1cm and -.7cm of X] (label) {\footnotesize\textbf{(a)}};

\draw[blue!70!black,->] (lambda) edge[bend right=20] (A1)
          (lambda) edge[bend left=20]  (B1)
          (lambda) edge[bend right=20] (A2)
          (lambda) edge[bend left=20] (B2)
;
\draw[->]
          (X) edge (A1)
          (X) edge[bend left=20] (A2)
          (Y) edge (B1)
          (Y) edge[bend right=20] (B2)
;       
\draw[->,red!70!black] 
          (A2) edge (K)
          (B2) edge (K)
;
\draw[o-o] (A1) edge (A2)
           (B1) edge (B2)
;
\end{tikzpicture}\\
\begin{tikzpicture}[thick, every node/.style={scale=1.2}]
\node[blue!70!black,text height=7pt,text depth=2pt] (lambda) { $\Lambda$};
\node[text height=7pt,text depth=2pt,below left = .7cm and -.2cm of lambda] (A2) {$\mathbf{A}^{(K)}$};
\node[text height=7pt,text depth=2pt, left = .7cm of A2] (A1) {$\mathbf{A}^{(X)}$};
\node[text height=7pt,text depth=2pt,below right = .7cm and -.2cm of lambda] (B2) {$\mathbf{B}^{(K)}$};
\node[text height=7pt,text depth=2pt, right = .7cm of B2] (B1) {$\mathbf{B}^{(Y)}$};
\node[text height=7pt,text depth=2pt,above = .3cm of A1] (X) {$\mathbf{X}$};
\node[text height=7pt,text depth=2pt,above = .3cm of B1] (Y) {$\mathbf{Y}$};
\node[draw,rectangle,red!70!black,text height=7pt,text depth=2pt,below = 1.8cm of lambda] (K) {$K$};
\node[text height=7pt,text depth=2pt,above left = 0.1cm and -.7cm of X] (label) {\footnotesize\textbf{(b)}};

\draw[blue!70!black,->] (lambda) edge[bend right=20] (A1)
          (lambda) edge[bend left=20]  (B1)
          (lambda) edge[bend right=20] (A2)
          (lambda) edge[bend left=20] (B2)
;
\draw[->]
          (X) edge (A1)
          (Y) edge (B1)
;       
\draw[->,red!70!black] 
          (A2) edge (K)
          (B2) edge (K)
;
\draw[->] (A2) edge (A1)
           (B2) edge (B1)
;
\end{tikzpicture}\\
\begin{tikzpicture}[thick, every node/.style={scale=1.2}]
\node[blue!70!black,text height=7pt,text depth=2pt] (lambda) { $\Lambda$};
\node[text height=7pt,text depth=2pt,below left = .7cm and -.2cm of lambda] (A2) {$\mathbf{A}^{(K)}$};
\node[text height=7pt,text depth=2pt, left = .7cm of A2] (A1) {$\mathbf{A}^{(X)}$};
\node[text height=7pt,text depth=2pt,below right = .7cm and -.2cm of lambda] (B2) {$\mathbf{B}^{(K)}$};
\node[text height=7pt,text depth=2pt, right = .7cm of B2] (B1) {$\mathbf{B}^{(Y)}$};
\node[text height=7pt,text depth=2pt,above = .3cm of A1] (X) {$\mathbf{X}$};
\node[text height=7pt,text depth=2pt,above = .3cm of B1] (Y) {$\mathbf{Y}$};
\node[draw,rectangle,red!70!black,text height=7pt,text depth=2pt,below = 1.8cm of lambda] (K) {$K$};
\node[text height=7pt,text depth=2pt,above left = 0.1cm and -.7cm of X] (label) {\footnotesize\textbf{(c)}};

\draw[blue!70!black,->] (lambda) edge[bend right=20] (A1)
          (lambda) edge[bend left=20]  (B1)
          (lambda) edge[bend right=20] (A2)
          (lambda) edge[bend left=20] (B2)
;
\draw[->]
          (X) edge (A1)
          (Y) edge (B1)
;       
\draw[->,red!70!black] 
          (A2) edge (K)
          (B2) edge (K)
;

\end{tikzpicture}
\caption{Causal diagrams to derive the causal structure to faithfully account for the fair sampling assumption (FSA). (a) We divide the measurement outcome variables $\mathbf{A}$ and $\mathbf{B}$ of a general bipartite Bell scenario into those that are used in the postselection decision ($\mathbf{A}^{(K)}$ and $\mathbf{B}^{(K)}$) and those that are not ($\mathbf{A}^{(X)}$ and $\mathbf{B}^{(Y)}$). Bidirected arrows indicate arbitrary causal influences between variables. (b) Causal diagram [restricting the one in (a)] that gives a faithful account of condition \textbf{CI}. (c) Causal diagram [restricting the one in (b)] that gives a faithful account of both conditions \textbf{CI} and \textbf{CII}, corresponding to a valid postselection and thus to the FSA.}
\label{fig:fairsampling_bipartite}
\end{figure}

To derive the causal structure required to show conditions $\textbf{CI}$ and $\textbf{CII}$ without fine-tuning, we first divide Alice's outcomes $\mathbf{A}$ into the outcomes $\mathbf{A}^{(K)}$ that are used to decide the postselection $K$, and the outcomes $\mathbf{A}^{(X)}$ that are not, see Fig.~\ref{fig:fairsampling_bipartite}(a). Similarly, we divide $\mathbf{B}$ into $\mathbf{B}^{(K)}$ and $\mathbf{B}^{(Y)}$. The bidirected arrows depict arbitrary causal influences, i.e., $A_i \multimapboth A_j$ includes $A_i\rightarrow A_j$, $A_i\leftarrow A_j$, or a hidden variable $\Gamma$ such that $A_i\leftarrow \Gamma \rightarrow A_j$ (a hidden common cause), and combinations of thereof. Note that a hidden common cause can be included in the LHV $\Lambda$. Now, if some $A_j\in \mathbf{A}^{(K)}$ is directly influenced by $\mathbf{X}$, there is an open path $\mathbf{X}\rightarrow A_j \leftarrow \Lambda$ because we condition on $K$, a descendant of the collider $A_j$ (we assume that the LHV $\Lambda$ influences all measurement outcomes). Thus, to preserve condition $\textbf{CI}$, $\mathbf{X}$ cannot have a direct influence on the group $\mathbf{A}^{(K)}$, and similarly for $\mathbf{Y}$ and $\mathbf{B}^{(K)}$. 
Next, if there was an influence $A_i\rightarrow A_j$ with $A_i\in \mathbf{A}^{(X)}$ and $A_j\in \mathbf{A}^{(K)}$, there would be an open path $X\rightarrow A_i \rightarrow A_j \leftarrow \Lambda$, in conflict with $\textbf{CI}$. Here we assume that $\mathbf{X}$ influences $A_i$ because, otherwise, $A_i$ would neither be useful for violating Bell inequalities\footnote{\label{footnote3}
To violate Bell inequalities, each party's setting must influence its measurement outcome. For instance, in the bipartite scenario with one setting and one measurement variable per party, assume that Alice's setting does not influence her outcome. Due to non-signalling, her setting cannot influence Bob's outcome, so one has
\begin{equation*}
    p_{ab|xy}=p_{ab|y}=p_{a|y}p_{b|ay}=p_{a}p_{b|ay}, 
\end{equation*}
where we have used the no-signalling principle $p_{a|y}=p_{a}$. This yields a LHV model for $p_{ab|xy}$: Defining $\Lambda$ to take the same values as $A$ with identical probabilities, $p_{\lambda}=p_a$ for $\lambda=a$, one has 
\begin{equation*}
    p_{ab|xy}=p_{a}p_{b|ay}=\sum_\lambda p_\lambda \delta_{a,\lambda}p_{b|\lambda,y}, 
\end{equation*}
where $\delta$ is the Kronecker symbol.
}, nor would it be useful to decide the postselection (because of $A_i\nrightarrow K$), so one can just discard the outcome $A_i$ from the analysis. We thus obtain Fig.~\ref{fig:fairsampling_bipartite}(b) which ensures that the condition $\textbf{CI}$ is fulfilled. For instance, for proving that $p_{\lambda|\mathbf{x}k}=p_{\lambda|k}$, note that the only path connecting $\mathbf{X}$ and $\Lambda$ passes through $\mathbf{A}^{(X)}$ that, being a collider, blocks the path.

Since there is no influence from $X$ to any $A_j\in \mathbf{A}^{(K)}$, the variables in $\mathbf{A}^{(K)}$ (and $\mathbf{B}^{(K)}$) are not useful to violate a Bell inequality$^\textrm{\ref{footnote3}}$. We thus only consider the variables $\mathbf{A}^{(X)}$ and $\mathbf{B}^{(Y)}$ as inputs to the Bell inequality. To show that the postselected statistics (describing correlations between $\mathbf{A}^{(X)}$ and $\mathbf{B}^{(Y)}$) can be described by a LHV model, it remains to show condition $\textbf{CII}$, i.e., $p_{a_ib_r|\mathbf{x}\mathbf{y}\lambda k} = p_{a_i|\mathbf{x}\lambda k}p_{b_r|\mathbf{y}\lambda k}$ for all $A_i\in \mathbf{A}^{(X)}$ and $B_r\in \mathbf{B}^{(Y)}$. If there were influences $A_j\rightarrow A_i$ and $B_s\rightarrow B_r$ for $A_i\in \mathbf{A}^{(X)}$, $A_j\in \mathbf{A}^{(K)}$, $B_r\in \mathbf{B}^{(Y)}$ and $B_s\in \mathbf{B}^{(K)}$, the path $A_i\leftarrow A_j \rightarrow K \leftarrow B_s \rightarrow B_r$ would be open, in conflict with $\textbf{CII}$. Excluding influences of the form $\mathbf{A}^{(K)} \rightarrow \mathbf{A}^{(X)}$ and $\mathbf{B}^{(K)} \rightarrow \mathbf{B}^{(Y)}$ ensures $\textbf{CII}$ because the only paths that connect $\mathbf{A}^{(X)}$ and $\mathbf{X}$ to $\mathbf{B}^{(Y)}$ and $\mathbf{Y}$ pass though $\Lambda$ and are blocked because $\Lambda$ is a non-collider that is conditioned on.
Thus, we conclude with Fig.~\ref{fig:fairsampling_bipartite}(c) which ensures both $\textbf{CI}$ and $\textbf{CII}$ (if restricted to variables in $\mathbf{A}^{(X)}$ and $\mathbf{B}^{(Y)}$). 

In summary, we have started from a general bipartite Bell scenario including a collective postselection and derived a necessary causal structure to faithfully describe the FSA, see Fig.~\ref{fig:fairsampling_bipartite}(c). This structure requires that each party must have at least one measurement variable ($\mathbf{A}^{(K)}$ and $\mathbf{B}^{(K)}$) that is used to decide the postselection and that is independent of the measurement settings, and at least one measurement variable ($\mathbf{A}^{(X)}$ and $\mathbf{B}^{(Y)}$) that is used as an input in the Bell inequality and that does not influence the postselection. The smallest realization of this structure is a Bell scenario where each party has a binary measurement setting ($X$ and $Y$), a binary measurement variable ($A_2$ and $B_2$) that dictates the postselection and a binary measurement variable ($A_1$ and $B_1$) that is used in the Bell inequality. In the standard use of the FSA to deal with the loss of particles, $A_2$ and $B_2$ correspond to the number of detected particles in the respective measurement stations, while the $A_1$ and $B_1$ correspond to the outcomes of, e.g., a polarization measurement of incoming photons\footnote{\label{footnote4} In this example, one could wonder about the meaning of $A_1$ if Alice does not detect a particle ($A_2=0$). Since this event will be discarded in the postselection, the value attributed to $A_1$ is not important. To be consistent with the assumption of Fig.~\ref{fig:fairsampling_bipartite}(c) ($A_2\nrightarrow A_1$) one could, e.g., flip a coin to set the value of $A_1$ in this case.}. We note however that, in general, the variables $\mathbf{A}^{(K)}$ and $\mathbf{B}^{(K)}$ not necessarily represent the number of detected particles, but the results hold for any collectively decided postselection.

We want to emphasize that the causal-diagram FSA is not only applicable to the standard scenario where one particle is sent to each party but there are detection and transmission losses, but also for certain postselection methods if the particles are generated in a superposition of their destinations~\cite{sciarrino2011,yurke1992b,yurke1992a}. In particular, demonstrations that a coincidence postselection is safe if the number of particles is conserved~\cite{blasiak2021,gebhart2021,gebhart2022} become unnecessary if one assumes the FSA. In other words, the FSA covers both a postselection due to inefficient detectors and transmission losses, and a postselection in ideal experiments due to a varying distribution of particles.

\begin{figure}[t]
    \centering
\begin{tikzpicture}[thick, every node/.style={scale=1.2}]
\node[blue!70!black,text height=7pt,text depth=2pt] (lambda) { $\Lambda$};
\node[draw,rectangle,text height=7pt,text depth=2pt,below left = .7cm and -.2cm of lambda] (A2) {$A_2$};
\node[text height=7pt,text depth=2pt, left = .7cm of A2] (A1) {$A_1$};
\node[draw,rectangle,text height=7pt,text depth=2pt,below right = .7cm and -.2cm of lambda] (B2) {$B_2$};
\node[text height=7pt,text depth=2pt, right = .7cm of B2] (B1) {$B_1$};
\node[text height=7pt,text depth=2pt,above = .3cm of A1] (X) {$X$};
\node[text height=7pt,text depth=2pt,above = .3cm of B1] (Y) {$Y$};

\draw[blue!70!black,->] (lambda) edge[bend right=20] (A1)
          (lambda) edge[bend left=20]  (B1)
          (lambda) edge[bend right=20] (A2)
          (lambda) edge[bend left=20] (B2)
;
\draw[->]
          (X) edge (A1)
          (Y) edge (B1)
          (A2) edge (A1)
          (B2) edge (B1)
;    
\end{tikzpicture}
    \caption{A possible causal diagram for the FSA if, for the postselection of the results, each party must receive a single particle. Here, $A_2$ ($B_2$) corresponds to the number of particles detected in Alice's (Bob's) measurement device.}
    \label{fig:fs_alternative}
\end{figure}
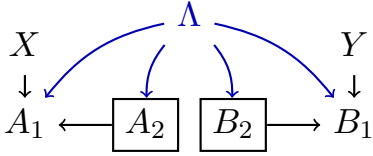

Above, we have derived a necessary structure of any bipartite causal diagram that faithfully describes the FSA for a general collective postselection. However, typical applications of the FSA are situations in which each measurement party needs to detect a single particle. Here, the variables $A_2$ and $B_2$ correspond to the number of detected particles in Alice's and Bob's experiment, respectively. This is a special case of a collective postselection, in which postselecting an event (denoted as $K=1$ above) is equivalent to a fixed combination of values for $A_2$ and $B_2$ ($A_2=B_2=1$). Thus, one can simply use a causal diagram with a conditioning on the variables $A_2$ and $B_2$ without introducing the postselection variable $K$. While influences such as $X\rightarrow A_2$ and $A_1\rightarrow A_2$ are still in conflict with condition \textbf{CI}, an influence of the form $A_2\rightarrow A_1$ can now be allowed. The corresponding causal diagram is shown in Fig.~\ref{fig:fs_alternative}. The causal diagram of Fig.~\ref{fig:fairsampling_bipartite}(c) (and its smallest realization) is more general though: An example of a collective postselection that cannot be modeled with Fig.~\ref{fig:fs_alternative} is when the parties postselect events for which $A_2=B_2$. Here, a postselected event does not imply fixed values of $A_2$ and $B_2$. 

To conclude this section, we want to mention that, while one cannot experimentally certify that the LHV is causally modelled by Fig.~\ref{fig:fairsampling_bipartite}(c) (in the same way that one cannot experimentally certify the FSA), our results highlight in which cases is it not possible to have a faithful causal account of the FSA. As discussed above, to account for the FSA, each party must have at least two separate measurement results, one that influences the postselection decision, and one that is used in the Bell inequality. This excludes setups for which the outcome used in the Bell inequality is also used to decide the postselection. An example is a 
setup proposed by Franson~\cite{franson1989} to create nonlocality from energy-time entanglement, which has been shown to admit LHV models that reproduce the observed statistics and the apparent Bell inequality violation even in the noiseless case~\cite{aerts1999,jogenfols2015,cabello2009,lima2010}. Here, each party has two different measurement outcomes, an early detection time and a late detection time. Even without particle losses, the statistics must be collectively postselected in order to violate a Bell inequality: Only those events are postselected for which both particles arrive either at the early or at the late detection time. Since the time of arrival is used both in the postselection and in the Bell inequality, there is no way to introduce two separate variables per party, e.g., $A_1$ and $A_2$, with the roles as above. Thus, a FSA for the noiseless version of the original Franson setup must rely on a fine-tuning in the causal diagram.

\subsection{Comparison to standard FSAs}

We now want to briefly compare the causal-diagram FSA to its different forms found in the literature. We emphasize that any of the following forms of the FSA corresponds to a fine-tuning condition on the original causal diagram of the Bell experiment (Fig.~\ref{fig:LHVbipartite}).  We first comment on the common (mis-)understanding that the FSA means that there is no observable influence of the measurement setting on the probability of detecting a particle, $p_{d|x}=p_d$, where $d$ represents Alice's detection of a particle. As shown in Ref.~\cite{berry2010} with a counter example, this assumption does not ensure a safe postselection. The original way of stating the FSA is that the postselected statistics should be a fair sample of (i.e., be identical to) the statistics that would have been obtained using perfect detectors~\cite{clauser1974}. This assumption is satisfied if $p_{d|x\lambda}=p_{d}$, i.e., if the probability of detecting a particle depends neither on the setting $X$ nor on the LHV $\Lambda$. This condition ensures a safe postselection but, as we have seen above, can be weakened: Assuming that $p_{d|x\lambda}=p_{d|\lambda}$ already provides a safe postselection~\cite{berry2010}. Here, the postselected ensemble may differ from the original one, $p_{\lambda|k}\neq p_\lambda$, but it still has the form of a LHV model, Eq.~\eqref{eq:LHV}. Assuming a causal-diagram representation without fine-tuning, the assumption that $X$ cannot influence the detection variable corresponds to the above causal diagrams of Fig.~\ref{fig:fairsampling_bipartite}(c) or Fig.~\ref{fig:fs_alternative}. 

Finally, we note that this FSA can be further weakened to the assumption that $p_{d|x\lambda}=\eta^{(d)}_x\eta^{(d)}_\lambda$~\cite{berry2010,orsucci2020}, i.e., the assumption that the detection efficiency depends on both the setting $X$ and the LHV $\Lambda$ but it factorizes. This factorization condition cannot be depicted in a causal diagram and, as the dependence implies that $A_2$ is influenced by both $X$ and $\Lambda$, it represents a fine-tuning of the causal influences.

\section{Fair sampling for genuine multipartite nonlocality}\label{sec:genuine}

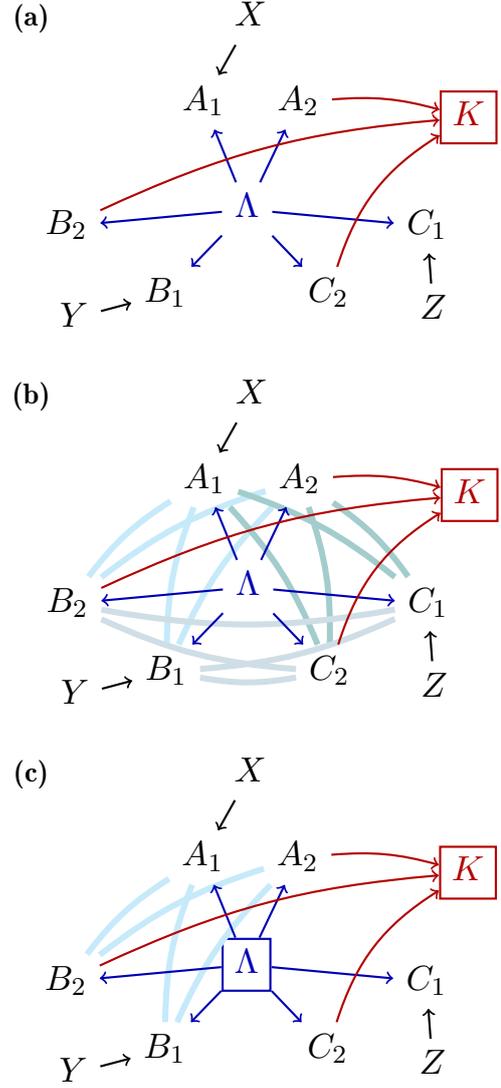
\begin{figure}[!ht]
\begin{center}
\begin{tikzpicture}[thick, every node/.style={scale=1.2}]
\centering 

\node[text height=7pt,text depth=2pt] (X) {$X$};
\node[text height=7pt,text depth=2pt,left=2.1cm of X] (lable) {{\footnotesize\textbf{(a)}}};
\node[text height=7pt,text depth=2pt,below left =.4cm and -.2cm of X] (A1) {$A_1$};
\node[text height=7pt,text depth=2pt,below right =.4cm and -.2cm of X] (A2) {$A_2$};
\node[blue!70!black,text height=7pt,text depth=2pt,below right =.7cm and -.2cm of A1] (Lambda) {$\Lambda$};
\node[text height=7pt,text depth=2pt,below left=-.5cm and 1.6cm of Lambda] (B2) {$B_2$};
\node[text height=7pt,text depth=2pt,below left=.4cm and .3cm of Lambda] (B1) {$B_1$};
\node[text height=7pt,text depth=2pt,below left=-.4cm and .4cm of B1] (Y) {$Y$};
\node[text height=7pt,text depth=2pt,below right=-.5cm and 1.6cm of Lambda] (C1) {$C_1$};
\node[text height=7pt,text depth=2pt,below right=.4cm and .3cm of Lambda] (C2) {$C_2$};
\node[text height=7pt,text depth=2pt,below right=-.5cm and .6cm of C2] (Z) {$Z$};
\node[draw,rectangle,red!70!black,text height=7pt,text depth=2pt,above right=.5cm and 2.2cm of Lambda] (K) {$K$};

\draw[blue!70!black,->] (Lambda) edge (A1)
          (Lambda) edge (A2)
          (Lambda) edge (B1)
          (Lambda) edge (B2)
          (Lambda) edge (C1)
          (Lambda) edge (C2)
;
\draw[->]
          (X) edge (A1)
          (Y) edge (B1)
          (Z) edge (C1)
;
\draw[->,red!70!black]
            (A2) edge[bend left=10] (K)
            (B2) edge[bend left=10] (K)
            (C2) edge[bend left=20] (K)
;
\end{tikzpicture}

\vspace{.3cm}

\begin{tikzpicture}[thick, every node/.style={scale=1.2}]
\centering 

\node[text height=7pt,text depth=2pt] (X) {$X$};
\node[text height=7pt,text depth=2pt,left=2.1cm of X] (lable) {{\footnotesize\textbf{(b)}}};
\node[text height=7pt,text depth=2pt,below left =.4cm and -.2cm of X] (A1) {$A_1$};
\node[text height=7pt,text depth=2pt,below right =.4cm and -.2cm of X] (A2) {$A_2$};
\node[blue!70!black,text height=7pt,text depth=2pt,below right =.7cm and -.2cm of A1] (Lambda) {$\Lambda$};
\node[text height=7pt,text depth=2pt,below left=-.5cm and 1.6cm of Lambda] (B2) {$B_2$};
\node[text height=7pt,text depth=2pt,below left=.4cm and .3cm of Lambda] (B1) {$B_1$};
\node[text height=7pt,text depth=2pt,below left=-.4cm and .4cm of B1] (Y) {$Y$};
\node[text height=7pt,text depth=2pt,below right=-.5cm and 1.6cm of Lambda] (C1) {$C_1$};
\node[text height=7pt,text depth=2pt,below right=.4cm and .3cm of Lambda] (C2) {$C_2$};
\node[text height=7pt,text depth=2pt,below right=-.5cm and .6cm of C2] (Z) {$Z$};
\node[draw,rectangle,red!70!black,text height=7pt,text depth=2pt,above right=.5cm and 2.2cm of Lambda] (K) {$K$};

\draw[-,line width=0.8mm,white!80!cyan]
          (A1) edge[bend right=10]  (B1)
          (A1) edge[bend right=10]  (B2)
          (A2) edge[bend right=10]  (B1)
          (A2) edge[bend right=10]  (B2)
;
\draw[-,line width=0.8mm,white!80!cyan!85!green!90!black]
          (A1) edge[bend left=10]  (C1)
          (A1) edge[bend left=10]  (C2)
          (A2) edge[bend left=10]  (C1)
          (A2) edge[bend left=10]  (C2)
;
\draw[-,line width=0.8mm,white!90!cyan!90!black]
          (B1) edge[bend right=10]  (C1)
          (B1) edge[bend right=10]  (C2)
          (B2) edge[bend right=10]  (C1)
          (B2) edge[bend right=10]  (C2)
;

\draw[blue!70!black,->] (Lambda) edge (A1)
          (Lambda) edge (A2)
          (Lambda) edge (B1)
          (Lambda) edge (B2)
          (Lambda) edge (C1)
          (Lambda) edge (C2)
;
\draw[->]
          (X) edge (A1)
          (Y) edge (B1)
          (Z) edge (C1)
;
\draw[->,red!70!black]
            (A2) edge[bend left=10] (K)
            (B2) edge[bend left=10] (K)
            (C2) edge[bend left=20] (K)
;
\end{tikzpicture}

\vspace{.3cm}

\begin{tikzpicture}[thick, every node/.style={scale=1.2}]
\centering 

\node[text height=7pt,text depth=2pt] (X) {$X$};
\node[text height=7pt,text depth=2pt,left=2.1cm of X] (lable) {{\footnotesize\textbf{(c)}}};
\node[text height=7pt,text depth=2pt,below left =.4cm and -.2cm of X] (A1) {$A_1$};
\node[text height=7pt,text depth=2pt,below right =.4cm and -.2cm of X] (A2) {$A_2$};
\node[blue!70!black,draw,rectangle,text height=7pt,text depth=2pt,below right =.7cm and -.2cm of A1] (Lambda) {$\Lambda$};
\node[text height=7pt,text depth=2pt,below left=-.5cm and 1.6cm of Lambda] (B2) {$B_2$};
\node[text height=7pt,text depth=2pt,below left=.4cm and .3cm of Lambda] (B1) {$B_1$};
\node[text height=7pt,text depth=2pt,below left=-.4cm and .4cm of B1] (Y) {$Y$};
\node[text height=7pt,text depth=2pt,below right=-.5cm and 1.6cm of Lambda] (C1) {$C_1$};
\node[text height=7pt,text depth=2pt,below right=.4cm and .3cm of Lambda] (C2) {$C_2$};
\node[text height=7pt,text depth=2pt,below right=-.5cm and .6cm of C2] (Z) {$Z$};
\node[draw,rectangle,red!70!black,text height=7pt,text depth=2pt,above right=.5cm and 2.2cm of Lambda] (K) {$K$};

\draw[-,line width=0.8mm,white!80!cyan]
          (A1) edge[bend right=10]  (B1)
          (A1) edge[bend right=10]  (B2)
          (A2) edge[bend right=10]  (B1)
          (A2) edge[bend right=10]  (B2)
;

\draw[blue!70!black,->] (Lambda) edge (A1)
          (Lambda) edge (A2)
          (Lambda) edge (B1)
          (Lambda) edge (B2)
          (Lambda) edge (C1)
          (Lambda) edge (C2)
;
\draw[->]
          (X) edge (A1)
          (Y) edge (B1)
          (Z) edge (C1)
;
\draw[->,red!70!black]
            (A2) edge[bend left=10] (K)
            (B2) edge[bend left=10] (K)
            (C2) edge[bend left=20] (K)
;
\end{tikzpicture}
\end{center}
   \caption{Examples of the FSA causal structure in the three-partite Bell scenario, for (a) a local hidden variable (LHV) model that corresponds to tests of multipartite nonlocality and (b,c) a hybrid local-nonlocal hidden variable model that corresponds to tests of genuine multipartite nonlocality. (c) In the hybrid model, when conditioning on a specific value of the LHV $\Lambda$, two of the three parties can share nonlocal quantum correlations (light blue lines) that are subject to the no-signalling fine-tuning condition.}
    \label{fig:threepartiteDAGs}
\end{figure}

The causal-diagram FSA of Fig.~\ref{fig:fairsampling_bipartite}(c) (or Fig.~\ref{fig:fs_alternative}) is also applicable in multipartite Bell experiments. For more than two measurement parties, there are different notions of nonlocality that can be demonstrated by a violation of the corresponding inequalities~\cite{svetlichny1987,mermin1990ineq,bancal2009,bancal2013,contreras2021}. For simplicity, we focus on the three-partite case, but the discussion holds for any number of parties. We thus include a third party Charlie who chooses a measurement setting $z$ and observes the measurement outcomes $(c_1,c_2)$, where for simplicity we only consider the smallest realization of the FSA causal structure including two measurement variables per party. First, one can assume a LHV model in the multipartite case similar to Eq.~\eqref{eq:LHV}, corresponding to a causal structure as shown in Fig.~\ref{fig:threepartiteDAGs}(a), where we included the FSA derived above. Using the LHV model, one can demonstrate inequalities that test multipartite nonlocality~\cite{mermin1990ineq}. The validity of the postselection, namely the conditions $p_{\lambda|xyzk}=p_{\lambda|k}$ and $p_{a_1b_1c_1|xyz\lambda k} = p_{a_1|x\lambda k}p_{b_1|y\lambda k}p_{c_1|z\lambda k}$, can be shown in exact analogy to the bipartite case above.

A second and stronger form of three-partite nonlocality is genuine three-partite nonlocality. Here, instead of assuming a LHV model, one allows for two of the three parties to share nonlocal quantum correlations, in a model that is called a hybrid local-nonlocal hidden variable model~\cite{svetlichny1987}\footnote{We note that, recently, several new definitions of (genuine) multipartite nonlocality have been proposed that we do not address in this work. These include network nonlocality~\cite{navascues2020}, broadcasting correlations~\cite{saha2015}, and nonlocality that is based on the resource theory of local operations and shared randomness (LOSR)~\cite{schmid2021,coiteux2021}. See also Ref.~\cite{chaves2017} for discussion of different classes of multipartite nonlocality.}. These quantum correlations, fulfilling the no-signalling principle~\cite{bancal2013}, cannot be depicted in a classical causal diagram without using fine-tuning conditions~\cite{wood2015,allen2017}. In Fig.~\ref{fig:threepartiteDAGs}, we indicate these correlations as light blue lines between the outcome variables, reminding that these influences are subject to the no-signalling principle. The hybrid model then dictates that, given a specific hidden variable $\lambda$ (i.e., when conditioning on $\Lambda$), there can only be nonlocal correlations between two of the parties, see Fig.~\ref{fig:threepartiteDAGs}(c). In contrast, when not conditioning on $\Lambda$, there can possibly exist nonlocal correlations between any pair of parties, see Fig.~\ref{fig:threepartiteDAGs}(b), where we use different colors to emphasize that only one pair of the parties can share nonlocal correlations at a time.

Hybrid local-nonlocal hidden variable models fulfill certain inequalities that test for genuine multipartite nonlocality~\cite{svetlichny1987}, and, similar to above, there are conditions on the postselected statistics that, if fulfilled, prove that a collective postselection is safe~\cite{gebhart2021}. One can directly show that, using the causal-diagram FSA as shown in Fig.~\ref{fig:threepartiteDAGs}, the conditions of a safe postselection are fulfilled. For instance, for the first condition, $p_{\lambda|xyzk}=p_{\lambda|k}$, we note that a path such as $X\rightarrow A_1\rightarrow B_2\rightarrow K\leftarrow C_2\leftarrow \Lambda$, that appears to be an open path since the collider $K$ is conditioned on, is blocked due to the no-signalling condition: Alice's measurement setting $X$ cannot influence Bob's measurement outcome $B_2$. The validity of the second condition of a factorization given a specific value of $\Lambda$, e.g., $p_{a_1b_1c_1|xyz\lambda k} = p_{a_1b_1|xy\lambda k}p_{c_1|z\lambda k}$, see Fig.~\ref{fig:threepartiteDAGs}(c), can again be seen by noting that any path that connects $C_1$ and $Z$ to the other parties passes through $\Lambda$, and is thus blocked because $\Lambda$ is a non-collider that is conditioned on. Thus we see that the FSA depicted in Fig.~\ref{fig:threepartiteDAGs} also suffices to validate a collective postselection for the demonstration of genuine multipartite nonlocality. 

Finally, as in the bipartite case, if each party needs to measure a single particle, the FSA can also be explained by a three-partite causal diagram similar to Fig.~\ref{fig:fs_alternative}. Here, $A_2$ represents the number detected particles in Alice's measurement, and one can allow for an influence of the form $A_2\rightarrow A_1$, and similarly for Bob and Charlie. Also this version of a fair-sampling causal diagram suffices to prove the conditions of a safe postselection for both multipartite nonlocality and genuine multipartite nonlocality. 


\section{Conclusions}\label{sec:conclusion}

We have discussed a causal explanation of the fair sampling assumption (FSA) that ensures that a collective postselection cannot create fake nonlocal correlations in Bell experiments. 
For this purpose, we have derived a causal structure that any causal model of a bipartite Bell scenario must possess to guarantee that the postselected statistics take the form of a local hidden variable (LHV) model, without requiring a fine-tuning of the causal influences. 
We have employed the framework of causal inference and $d$-separation rules~\cite{pearl2009} as a mediator between a causal structure and the implied relations of conditional independence between the variables. 
Our results clarify what one really assumes if one uses the FSA, and while the corresponding causal structure is not experimentally certifiable (similar to the FSA itself), our results demonstrate that, in certain Bell scenarios, there is no faithful causal account of the FSA.
The derived FSA causal structure yields an easy and intuitive explanation of the FSA and can be used to understand different forms of the FSA in the literature.
Furthermore, besides standard Bell scenarios with a non-ideal detection and particle losses, we have demonstrated that the causal-diagram FSA can also be applied in noise-free scenarios where the statistics must be postselected because the particles are randomly distributed between the parties~\cite{yurke1992b,yurke1992a,gebhart2021,gebhart2022}.
Finally, we have shown that the FSA is also applicable for a collective postselection in demonstrations of multipartite nonlocality and genuine multipartite nonlocality.

\section*{Acknowledgments}
This work was supported by the European Commission through the H2020 QuantERA ERA-NET Cofund in Quantum Technologies project “MENTA”



\end{document}